\documentclass[aps,pra,showkeys,twocolumn,superscriptaddress]{revtex4-2}
\usepackage{amsmath, amssymb}
\usepackage{mathrsfs}
\usepackage{array}
\usepackage{booktabs}
\usepackage{tabu}
\usepackage{dcolumn}
\usepackage{amsmath}
\usepackage{amsfonts}
\usepackage{float}
\usepackage{amssymb}
\usepackage{graphicx,color}
\usepackage[colorlinks={true}]{hyperref}
\hypersetup{colorlinks=true,linkcolor=red,citecolor=magenta,urlcolor=blue}
\usepackage{graphicx}
\usepackage{subfigure}
\usepackage{graphicx}
\usepackage{dcolumn}
\usepackage{bm}
\usepackage{pstricks}
\usepackage{braket}
\usepackage{orcidlink}

\def\be{\begin{equation}}
  \def\ee{\end{equation}}
\def\bea{\begin{eqnarray}}
\def\eea{\end{eqnarray}}
\def\f{\frac}
\def\n{\nonumber}
\def\l{\label}
\def\p{\phi}
\def\o{\over}
\def\R{\mathcal{R}}
\def\pa{\partial}
\def\om{\omega}
\def\na{\nabla}
\def\P{\Phi}
\begin{document}

\title{Performance of a Superconducting Quantum Battery}

\author{Samira Elghaayda~\!\!\orcidlink{0000-0002-6655-0465}}
\affiliation{ Laboratory of High Energy Physics and Condensed Matter, Department of Physics,\\ Faculty of Sciences of Aïn Chock, Hassan II University,\\ P.O. Box 5366 Maarif, Casablanca 20100, Morocco.}

\author{Asad Ali~\!\!\orcidlink{0000-0001-9243-417X}} \email{asal68826@hbku.edu.qa}
\affiliation{Qatar Centre for Quantum Computing, College of Science and Engineering, Hamad Bin Khalifa University, Doha, Qatar}

\author{Saif Al-Kuwari~\!\!\orcidlink{0000-0002-4402-7710}}
\email{smalkuwari@hbku.edu.qa}
\affiliation{Qatar Centre for Quantum Computing, College of Science and Engineering, Hamad Bin Khalifa University, Doha, Qatar}

\author{Artur Czerwinski~\!\!\orcidlink{0000-0003-0625-8339}}
\affiliation{Institute of Physics, Faculty of Physics, Astronomy and Informatics, Nicolaus Copernicus University in Torun,
ul. Grudziadzka 5, 87-100 Torun, Poland}

\author{Mostafa Mansour~\!\!\orcidlink{0000-0003-0821-0582}}
\affiliation{ Laboratory of High Energy Physics and Condensed Matter, Department of Physics,\\ Faculty of Sciences of Aïn Chock, Hassan II University,\\ P.O. Box 5366 Maarif, Casablanca 20100, Morocco.}

\author{Saeed Haddadi~\!\!\orcidlink{0000-0002-1596-0763}} \email{haddadi@ipm.ir}
\affiliation{School of Particles and Accelerators, Institute for Research in Fundamental Sciences (IPM), P.O. Box 19395-5531, Tehran, Iran}

\date{\today}
\def\be{\begin{equation}}
  \def\ee{\end{equation}}
\def\bea{\begin{eqnarray}}
\def\eea{\end{eqnarray}}
\def\f{\frac}
\def\n{\nonumber}
\def\l{\label}
\def\p{\phi}
\def\o{\over}
\def\R{\mathcal{R}}
\def\pa{\partial}
\def\om{\omega}
\def\na{\nabla}
\def\P{$\Phi$}

\begin{abstract}
Finding a quantum battery model that demonstrates a quantum advantage while remaining feasible for experimental production is a considerable challenge. Here, a superconducting quantum battery (SQB) model that exhibits such an advantage is introduced. The model consists of two coupled superconducting qubits that interact during the unitary charging process while remaining in equilibrium with a thermal reservoir. First, the model is described, evidence of the quantum advantage is provided, and then the fabrication process of the battery is discussed using superconducting qubits. Analytical expressions for the ergotropy, instantaneous power, and capacity of the SQB, as well as their connection to quantum coherence are derived. It is demonstrated that leveraging the collective effects of Josephson energies and the coupling energy between qubits allows for optimization, resulting in improved energy redistribution and a significant enhancement in charging efficiency. This work highlights the complexities of tuning system parameters, which increase the potential for work extraction from the SQB, providing a deeper understanding of the charging mechanisms involved. These findings can be applied to superconducting quantum circuit battery architectures, underscoring the feasibility of efficient energy storage in these systems. These results pave the way for proposals of new superconducting devices, emphasizing their potential for efficient energy storage.
\end{abstract}


\maketitle

\section{Introduction}\label{sec0}
Over the last twenty years, we have witnessed significant progress in quantum technologies, ranging from fundamental physics to commercial applications. This advancement has led to important developments in energy transfer mechanisms within the quantum realm and the role of quantum resources in microscopic devices \cite{Crawford2021, AUMASSON20178, Radanliev2024, Bass2023QuantumSF, Degen2016QuantumS,zhang2023quantum,arjmandi2022enhancing}. By focusing on these advancements, one can see various physical implementations emerging as leading technologies, such as quantum batteries \cite{campaioli2024colloquium, quach2023quantum, quach2020organic, alicki2013entanglement} and superconducting circuits \cite{clerk2020hybrid, xiang2013hybrid, kjaergaard2020superconducting}. While it is still too early to determine which of these will form the foundation of tomorrow's quantum batteries \cite{campaioli2024colloquium, quach2023quantum, quach2020organic, alicki2013entanglement}, superconducting qubits are arguably at the forefront \cite{hu2022optimal,PhysRevA.107.023725,benzahra2023quantum}. Research in this area has progressed beyond academia and is now attracting significant interest from major technology companies, including
International Business Machines(IBM) corporation, underscoring its impact on the field.

Quantum batteries can serve as stationary systems that store energy in their excited states for later use, or as dynamic mediums that transfer energy to other systems. Traditional energy storage solutions, such as chemical batteries, face several challenges, such as slow charging, short lifespans, and harmful environmental impacts \cite{dell2001understanding}. Despite some fundamental limitations \cite{Gyhm2024}, quantum batteries are designed to tackle these challenges by leveraging superabsorption and superradiant states, which allow multiple quantum systems to absorb energy more efficiently than classical systems, allowing faster charging rates with minimal energy loss during charging and discharging \cite{ferraro2018high, Yang2024}. Progress has been made in developing these tiny devices experimentally \cite{ahmadi2024nonreciprocal}.

Superconducting qubits made from Josephson junctions are promising for quantum sensing, hybrid systems, and large-scale quantum computing architectures \cite{clerk2020hybrid, xiang2013hybrid,kjaergaard2020superconducting}. Current qubits operate below 10 GHz using standard cryogenic microwave equipment. Higher frequencies broaden the range of energies available for quantum experiments, enabling connections to a wider array of signals and quantum emitters, and allowing for higher operating temperatures. This scalability method leverages enhanced cooling power with even a slight increase in temperature  \cite{pobell2007matter}. Thermally robust qubits could reduce hardware demands for microwave quantum interconnects \cite{pechal2017millimeter, magnard2020microwave}, integrate with superconducting digital logic \cite{liu2023single}, and help manage heat from growing numbers of qubit control lines \cite{krinner2019engineering} as superconducting quantum processors scale to hundreds of qubits \cite{wu2021strong}. Resilient qubits in hybrid quantum systems could simplify experiments and improve performance \cite{clerk2020hybrid, xiang2013hybrid}.

Optimizing charging processes is a core objective in quantum battery research. Effective quantum control techniques are essential for enabling faster and more reliable charging. Recently, many researchers studying the charging of quantum batteries have shown that quantum coherence and entanglement significantly enhance charging speed and energy storage efficiency \cite{kamin2020entanglement, shi2022entanglement}. Furthermore, research on the scalability of quantum battery performance with system size reveals that collective quantum effects enable enhanced energy extraction beyond classical capabilities \cite{andolina2018charger,rossini2019many, yao2022optimal, dou2022cavity,PhysRevE.106.054107,PhysRevE.109.064103, mojaveri2024extracting,ali2024ergotropy,wangAQT2025,HaddadiQB2024}. However, the charging protocols often rely on specific initial battery states. This dependence on the state of the quantum battery remains relatively unexplored. In this work, along these lines, we address and highlight a superconducting quantum battery (SQB)  initialized in a thermal state. This battery is constructed from superconducting charge qubits with symmetric Josephson junction energies. These qubits are capacitively coupled and shunted with a large capacitance to suppress charge noise. Known as Transmon qubits, they function as artificial atoms, allowing us to harness their Josephson energies, particularly the mutual coupling energy between qubits, to drive ergotropy production and enhance both energy and ergotropy. SQBs represent a cutting-edge paradigm in quantum energy storage \cite{hu2022optimal}, aiming to exploit quantum correlations and superposition to achieve charging capabilities that surpass those of conventional electrochemical batteries in efficiency and performance.

\subsection{Contribution and Motivation}
The quest for efficient quantum battery models that exhibit quantum advantages is a challenge in advancing quantum technology. Superconducting qubits, with their established role in quantum computing, offer a promising platform for realizing quantum batteries. However, developing a feasible design that optimizes energy transfer requires careful consideration of coherence and interaction dynamics. In this work, we propose an SQB model based on two coupled superconducting qubits. This model demonstrates a significant quantum advantage in energy storage and extraction while remaining experimentally feasible. By analyzing the SQB's ergotropy, power, and capacity, we identify key factors, such as coherence and qubit coupling, that maximize work extraction, illustrating how collective effects from Josephson energies and inter-qubit coupling enhance charging efficiency. This SQB model connects the quantum properties of the system with practical considerations for device fabrication, laying a foundation for developing high-performance energy storage in superconducting quantum circuits.

Specifically, we focus on a closed SQB, where the charging process is inherently unitary and free from decoherence and dissipation effects. This establishes a theoretical benchmark for the optimal performance of an SQB under ideal coherent dynamics. Although our model includes a thermal reservoir, it does not induce dissipation during the charging process; the reservoir only defines the initial equilibrium state of the qubits, ensuring that the system remains effectively closed throughout the unitary evolution. In practical superconducting architectures, qubits are subject to noise sources such as charge noise, flux noise, and relaxation effects. While these are not explicitly considered in our analysis, they represent an important aspect for future investigations into the robustness of the observed quantum advantage under realistic conditions. This work could be further developed in two main directions. First, we propose a universal method for fast charging of an SQB connected to a driven quantum charger system through controlled pure dephasing of the charger. Second, we present a feasible scheme for an SQB that can implement the proposed cycle and provide typical experimental values for the circuit elements.

In particular, the relationship between decoherence time and charging time is crucial for the efficiency of SQB. If the charging time is significantly shorter than the decoherence time, the system can achieve near-optimal performance before noise-induced energy losses become significant. Conversely, if decoherence effects occur on comparable or shorter timescales, the benefits of quantum coherence and entanglement may diminish. Recent studies on superconducting qubit coherence times suggest that state-of-the-art devices exhibit coherence times ranging from tens to hundreds of microseconds \cite{kjaergaard2020superconducting}, which should be considered when designing practical charging protocols.

Although the current work does not explicitly model decoherence, our results will provide a theoretical benchmark for the optimal performance of an SQB in the absence of noise. Future extensions could incorporate open-system effects using techniques such as Lindblad master equations or stochastic noise modeling to assess the SQB's robustness against decoherence \cite{rivas2012open,czerwinski2022dynamics}. Recent advancements in quantum error mitigation and dynamical decoupling techniques \cite{PRXQuantum.2.010339} offer promising methods to extend coherence times and enhance the viability of SQBs in real-world implementations.

\subsection{Organization}
In this work, we focus on the role of coherence in ergotropic work extraction. We begin by briefly reviewing the relevant model of two superconducting charge qubits (SCQs) in Sec. \ref{sec1}. The remainder of this section discusses the fabrication of the SQB within this framework and presents the details of the charging process. It also includes fundamental performance indicators for SQBs, such as ergotropy, power, capacity, and coherence. In Sec. \ref{sec2}, we qualitatively describe the results and insights gained regarding the performance of the SQB.  A discussion on the scalability of SQBs is provided in Sec. \ref{sec3}. Experimental feasibility and optimization strategies for SQBs are presented in Sec. \ref{sec4}.
We summarize our findings and discuss potential future research directions in the final section, Sec. \ref{sec5}.

\section{Model and performance of SQB \label{sec1}}
The current model consists of two single Cooper-pair box charge qubits coupled through a fixed capacitor \cite{shaw2009characterization,paraoanu2006microwave,li2008entanglement,tian2011entanglement}. Superconducting materials in the model exhibit a suppressed transition temperature \( T_c \), which is carefully adjusted by selecting suitable materials, typically achieving values in the milli-Kelvin temperature range to facilitate efficient and scalable superconducting circuits. A representative example is a superconducting trilayer made of Al/Ti/Au, with respective layer thicknesses of \( 300 \, \text{Å} \), \( 200 \, \text{Å} \), and \( 200 \, \text{Å} \), and a transition temperature of \( T_c = 450 \, \text{mK} \) \cite{shaw2009characterization,paraoanu2006microwave,li2008entanglement,tian2011entanglement}. The Hamiltonian of a two-superconducting-qubit system can be expressed by \cite{shaw2009characterization}

\begin{align}
  \mathcal{H}&= -\frac{1}{2} \bigg\{ \left[ 4 \xi_{c1} \left( \frac{1}{2} - n_{g1} \right) + 2 \xi_c \left( \frac{1}{2} - n_{g2} \right) \right] \sigma_{z1} \notag \\
	&\quad + \left[4 \xi_{c2} \left( \frac{1}{2} - n_{g2} \right) + 2 \xi_c \left( \frac{1}{2} - n_{g1} \right) \right] \sigma_{z2} \notag \\
	&\quad + \xi_{1} \sigma_{x1} + \xi_{2} \sigma_{x2} - 2 \xi_c \sigma_{zz} \bigg\},
	\label{eq7}
\end{align}
where $\xi_{c1}$ and $\xi_{c2}$ are the charging energies,  $\xi_{1}$ and $\xi_{2}$ are the Josephson energies, and $\xi_c$ is the mutual coupling energy between the two qubits. Besides, $\sigma_{x(z)1} = \sigma_{x(z)} \otimes I$, $\sigma_{x(z)2} = I \otimes \sigma_{x(z)}$, and $\sigma_{zz} = \sigma_z \otimes \sigma_z$, with $\sigma_{x,z}$ being Pauli matrices and $I$ is the identity matrix. Also, $n_{gj} = \frac{C_{gj} V_{gj}}{2e}$ is the normalized qubit gate charge, where $C_{gj}$ and $V_{gj}$ are the control gate capacitance and voltage, respectively.

In our analysis, we restrict our calculations to the degeneracy point ($ n_{g1} = n_{g2} = 1/2 $), as this condition minimizes charge noise sensitivity and increases the coherence properties of the superconducting qubits \cite{paraoanu2006microwave}. This choice is a widely adopted strategy in superconducting qubit studies, where operating at the degeneracy point allows for more stable quantum coherence, reducing the impact of environmental fluctuations \cite{vion2002manipulating}. By focusing on this regime, we ensure that the system remains robust against low-frequency charge noise, which is one of the dominant decoherence mechanisms in superconducting circuits \cite{Nakamura1999}. However, we acknowledge the importance of investigating the system’s behavior at different operating points, as deviations from the degeneracy point may introduce charge dispersion effects that influence qubit dynamics. Future work will extend this model to analyze how variations in $n_{g1} $ and $ n_{g2} $ affect quantum correlations and coherence, providing a more comprehensive understanding of the system’s performance under realistic operating conditions. Such an extension is crucial for assessing the practical feasibility of implementing superconducting qubits in scalable quantum technologies, where external perturbations and fabrication imperfections may lead to operation outside the ideal degeneracy condition \cite{PhysRevA.76.042319}. Furthermore, a broader analysis will help characterize the interplay between charge noise and qubit interactions, offering insights into potential mitigation strategies for improving qubit fidelity in experimental implementations.

Under this condition, the Hamiltonian \eqref{eq7} simplifies to

\begin{align}
    \mathcal{H_{B}} =-\frac{1}{2} \left[ \xi_{1} \sigma_{x1}  + \xi_{2} \sigma_{x2}- \xi_{c} \sigma_{zz} \right].
\end{align}
The above Hamiltonian in matrix form yields

\begin{equation}\label{hm}
\mathcal{H_{B}} = \frac{1}{2}\left(
\begin{array}{cccc}
 2\xi _c & -\xi _2 & -\xi _1 & 0 \\
 -\xi _2 & -2\xi _c & 0 & -\xi _1 \\
 -\xi _1 & 0 & -2\xi _c & -\xi _2 \\
 0 & -\xi _1 & -\xi _2 & 2\xi _c \\
\end{array}
\right).
\end{equation}
At \(t = 0\), if the system is fully discharged, its initial state can be the ground state \(\mathcal{R}(0) = |00\rangle \langle 00|\) at absolute zero or Gibbs state at any finite temperature. The effectiveness of a quantum battery depends on a robust charging process; however, these processes are often sensitive to the initial state of the battery. We assume the Gibbs state as the initial state of quantum battery instead of ground state as it is practical experimentally to deal with finite temperature quantum battery than quantum battery at ground state. Therefore, for a system at finite temperature \(T\), we introduce the initial state described by the Gibbs thermal state:
\begin{equation}
\mathcal{R}_{th}(0) = \frac{e^{-\beta \mathcal{H_{B}}}}{\mathcal{Z}},
\end{equation}
where \(\beta = \frac{1}{k_B T}\) is the inverse temperature (set the Boltzmann constant $k_B=1$) and $\mathcal{Z} = \text{Tr}(e^{-\beta \mathcal{H_{B}}})$ is the partition function. This thermal state represents the probability distribution of the energy eigenstates of the system at the specified temperature. Given the structure of the Hamiltonian, the thermal state can be expressed as follows:

\begin{equation}
	\mathcal{R}_{th}(0) = \frac{1}{\mathcal{Z}}\left(\begin{array}{cccc}
		\mathcal{R}_{th_{11}} & \mathcal{R}_{th_{12}} & \mathcal{R}_{th_{13}} & \mathcal{R}_{th_{14}} \\
		\mathcal{R}_{th_{21}} & \mathcal{R}_{th_{22}} & \mathcal{R}_{th_{23}} & \mathcal{R}_{th_{24}}  \\
		\mathcal{R}_{th_{31}}  & \mathcal{R}_{th_{32}} & \mathcal{R}_{th_{33}} & \mathcal{R}_{th_{34}} \\
		\mathcal{R}_{th_{41}} & \mathcal{R}_{th_{42}}  & \mathcal{R}_{th_{43}} & \mathcal{R}_{th_{44}}
	\end{array}\right). \label{eq:5}
\end{equation}
The above matrix elements are reported in Appendix \ref{AppA}.

\begin{figure}[t]
    \centering
    \includegraphics[width=.999\columnwidth]{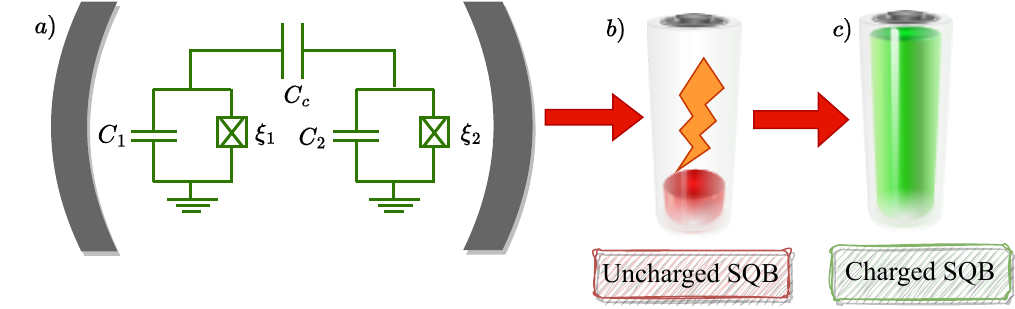}
    \caption{ Schematic circuit of the proposed SQB charging protocol. The system is initially a) prepared in a Gibbs state, consisting of two superconducting qubits at thermal equilibrium. These qubits correspond to the two Josephson junctions and represent the first and second Cooper pair boxes, respectively. Charging of the SQB occurs under a local field pulse $\Omega(t)$, which provides the necessary energy. Panels b) and c) represent the uncharged and charged states of the battery, respectively.}
    \label{f0}
    \end{figure}
Now, we can evaluate the performance of our SQB by charging it via the Pauli-X gate-based charging Hamiltonian, which is discussed in the following subsections. The model consists of two coupled superconducting qubits that interact during the unitary charging process while remaining in equilibrium with a thermal reservoir (see Fig. \ref{f0}).

\subsection{Charging closed SQB with Pauli-X gate}
We consider the charging process of a SQB in a closed
quantum system, where charging protocols are necessarily unitary. First, we note that for almost every density matrix, there exists a unitary map that increases its energy. This can be understood by recognizing that the problem of charging with respect to \( \mathcal{H_{B}} \) is equivalent to discharging with respect to \( -\mathcal{H_{B}}\). Therefore, the only states that cannot be charged using a unitary protocol are those whose ergotropy \cite{allahverdyan2004maximal} with respect to \( -\mathcal{H_{B}} \) is zero. From Refs. \cite{lenard1978thermodynamical, pusz1978passive, salvia2021distribution}, we conclude that the only states that cannot be charged by any unitary operation are those that are diagonal in the eigenbasis of \(\mathcal{H_{B}} \) with eigenvalues that are non-decreasing in energy.

The SQB can be charged by applying a local field pulse in the $x$-direction with pulse strength \(\Omega = \Omega(t)\). This field transfers energy to charge the SQB. The expression for the charging Hamiltonian \(\mathcal{H}_{c_x}\) in the $x$-direction is

\begin{equation}
   \mathcal{H}_{c_x} = \Omega (\sigma_{x} \otimes \mathbb{I}_{2} + \mathbb{I}_{2} \otimes \sigma_{x}).
\end{equation}
In practice, the operation of the X-gate in superconducting qubits is achieved by applying a microwave pulse that induces a transition between the qubit's ground state \( |0\rangle \) and excited state \( |1\rangle \). This pulse, resonant with the qubit's transition frequency, is shaped to effect a \(\pi\)-radian rotation around the $x$-axis of the Bloch sphere, thus flipping the qubit's state. The pulse duration, amplitude, and phase are calibrated to ensure high-fidelity state inversion. Typically, a quantum battery is charged by delivering uniform pulses to all quantum cells. During the charging process, once the quantum battery reaches its peak value, \(t \to \tau\), the $x$-direction pulse is turned off to prevent the quantum battery from returning to its initial state. In a closed system, the charging process can be implemented using the unitary operation
\begin{equation}\label{uni}
\mathcal{U}_{\mathcal{X}}(t)=\exp[-i\mathcal{H}_{c_x} t],
\end{equation}
which drives the qubit system from its initial state. The unitary operator for X-gate charging can be explicitly expressed as
\begin{equation}\label{eq:8}
\mathcal{U}_{\mathcal{X}}(t) = \left(
\begin{array}{cccc}
 \textrm{a} & \textrm{c} & \textrm{c} & \textrm{b} \\
 \textrm{c} & \textrm{a} & \textrm{b} & \textrm{c} \\
 \textrm{c} & \textrm{b} & \textrm{a} & \textrm{c} \\
 \textrm{b} & \textrm{c} & \textrm{c} & \textrm{a} \\
\end{array}
\right),
\end{equation}
here, $\textrm{a} = \cos^2(\tau)$, $\textrm{b} = -\sin^2(\tau)$, and $\textrm{c} = -i \sin(\tau) \cos(\tau)$ with $\tau=\Omega t$.


\subsection{Ergotropy, power, capacity, and quantum coherence}

To gain more insights into the working of our system, we analyze our Hamiltonian through spectral decomposition given by

\begin{equation}\label{spec}
\mathcal{H_{B}} = \sum_{i=1}^{4} \epsilon_i |\Delta_i\rangle\langle\Delta_i|, \text{ with } \epsilon_{i+1} \geq \epsilon_i.
\end{equation}
In this context, $\epsilon_i$'s represent the eigenvalues, and $|\Delta_i\rangle$'s denote the corresponding eigenstates. This representation is helpful for assessing energy changes and work extraction potential in cyclic unitary processes, which suggests that the Hamiltonian of the quantum battery must be the same at the end as it was in its initial state.

The state $\mathcal{R}$ is expressed as:
\begin{equation}\label{state}
\mathcal{R} = \sum_{j=1}^{4} \lambda_j |e_j\rangle\langle e_j|, \text{ where } \lambda_{j+1} \leq \lambda_j.
\end{equation}

The goal is to convert the system from the state $\mathcal{R}$ to a state with lower internal energy to extract the energy difference. After extracting the maximum amount of work in a cyclic and unitary process, the system reaches a passive state $\pi$ \cite{ghosh2020enhancement,binder2015quantacell,pusz1978passive,allahverdyan2004maximal}. This passive state is diagonal in the Hamiltonian's eigenbasis and has a decreasing population with increasing energy levels. We define this passive state as $\pi = \mathcal{R}_{\text{th}}$, a Gibbs thermal state.
To gain a deeper understanding of energy extraction efficiency from quantum states, it is important to study the maximum amount of work that can be extracted from a quantum system through a cyclic unitary process, a concept known as ergotropy. This process involves transitioning the system from its initial state to a thermal state. Ergotropy is defined as:

\begin{equation}\label{ergo}
\mathcal{E}:= \sum_{m,n} \lambda_m \epsilon_n \left[|\langle \Delta_n | e_m \rangle|^2 - \delta_{mn}\right],
\end{equation}
where $\delta_{mn}$ is the Kronecker delta.
Alternatively,
\begin{equation}\label{ergo2}
\mathcal{E}:= \text{Tr}[(\mathcal{R} - \mathcal{R}_{\text{th}})\mathcal{H_{B}}].
\end{equation}

The figure of merit we use to analyze this model is ergotropy and average power of SQB. Note that the instantaneous power of a quantum battery is defined as $\mathcal{P}=d\mathcal{E}/dt$.

If the final state is another non-passive state $\Pi$, the work done is:
\begin{equation}
\mathcal{W} = \text{Tr}[(\mathcal{R} - \Pi)\mathcal{H_{B}}].
\end{equation}

Furthermore, the capacity of a quantum battery, denoted as $\mathcal{K}$, serves as an effective performance indicator \cite{yang2023battery}. This metric allows us to gain meaningful insights into the quantum battery without the need to solve intricate dynamical equations. The capacity of a quantum battery is characterized as:
\begin{equation}
\mathcal{K} = \text{Tr}[\mathcal{H_{B}} \hat{\rho}_{\uparrow}] - \text{Tr}[\mathcal{H_{B}} \hat{\rho}_{\downarrow}],
\label{eq32}
\end{equation}
where $\hat{\rho}_{\downarrow} = |0^{\otimes N}\rangle\langle 0^{\otimes N}|$ and $\hat{\rho}_{\uparrow} = |1^{\otimes N}\rangle\langle 1^{\otimes N}|$ represent the ground and excited states of the $N$-partite quantum battery, respectively. For our case of two qubits, we have \(N=2\). This measure directly quantifies the energy gap between the maximum and minimum energy states of the quantum battery. Its calculation is straightforward and does not require time-dependent optimization, making it suitable for both open and closed quantum batteries. Therefore, \(\mathcal{K}\) serves as a practical and efficient metric for evaluating quantum battery performance.


In order to keep track of the degree of the quantum coherence existing inside SQB while the work is being extracted in the form of ergotropy, we evaluate $l_1$-norm of quantum coherence \cite{baumgratz2014quantifying}. Specifically, we consider the following density matrix
\begin{equation}\label{RX}
\mathcal{R_X} (t)=\mathcal{U}_{\mathcal{X}}(t) \mathcal{R}_{th}(0) \mathcal{U}_{\mathcal{X}}(t)^\dagger,
\end{equation}
where $\mathcal{R}_{th}(0)$ and $\mathcal{U}_{\mathcal{X}}(t)$ are defined in Eqs. \eqref{eq:5} and \eqref{eq:8}, respectively.
The explicit form of the $\mathcal{R_X}(t)
$ is given in the Appendix \ref{appB}. Hence, one can evaluate the $l_1$-norm of quantum coherence as follows
\begin{equation}
\mathcal{C}_{l_1}(\mathcal{R_X} (t)) = \sum_{i \neq j} |\langle i | \mathcal{R_X} (t) | j \rangle|.
\label{eq31}
\end{equation}

Note that we provide the closed-form expressions for the performance metrics in Appendix \ref{appC}.

\begin{figure*}[t]
			\centering
			\subfigure[]{\label{figure1a}\includegraphics[scale=0.5]{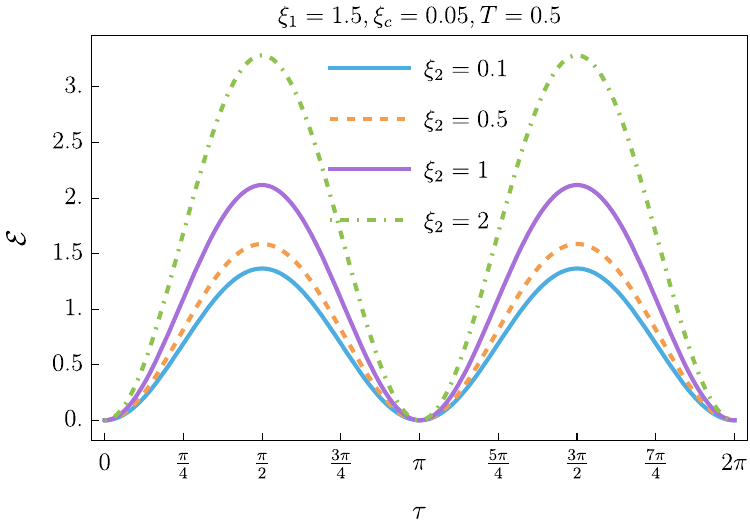}}
			\subfigure[]{\label{figure1b}\includegraphics[scale=0.5]{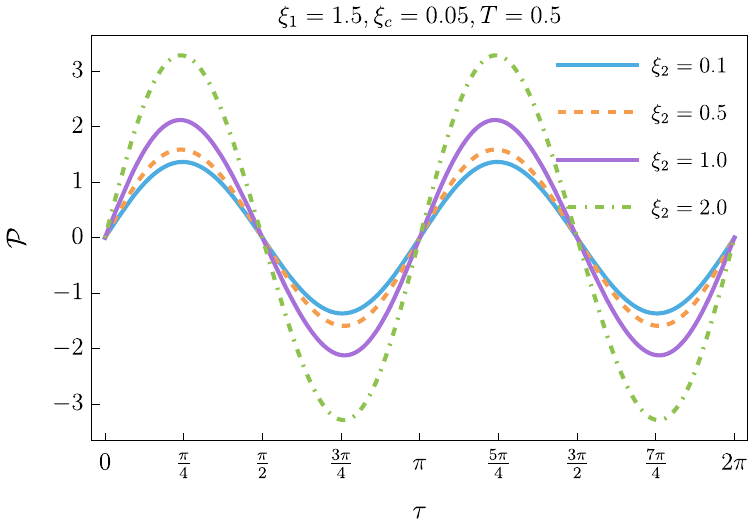}}
   \subfigure[]{\label{figure1c}\includegraphics[scale=0.5]{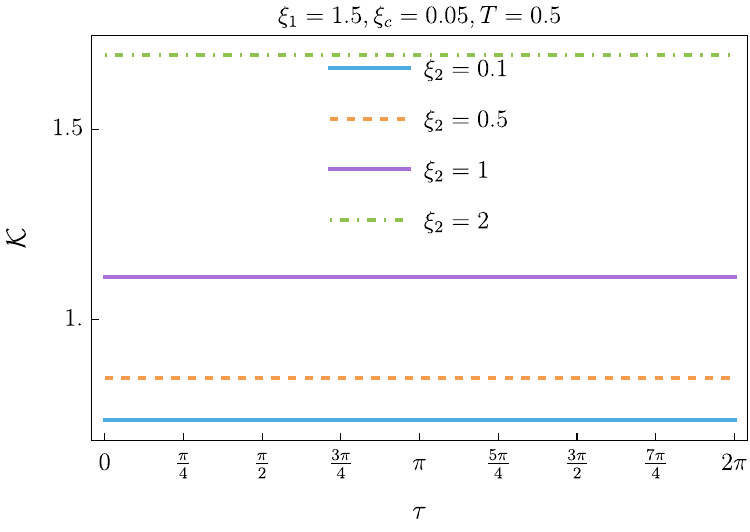}}
			\subfigure[]{\label{figure1d}\includegraphics[scale=0.5]{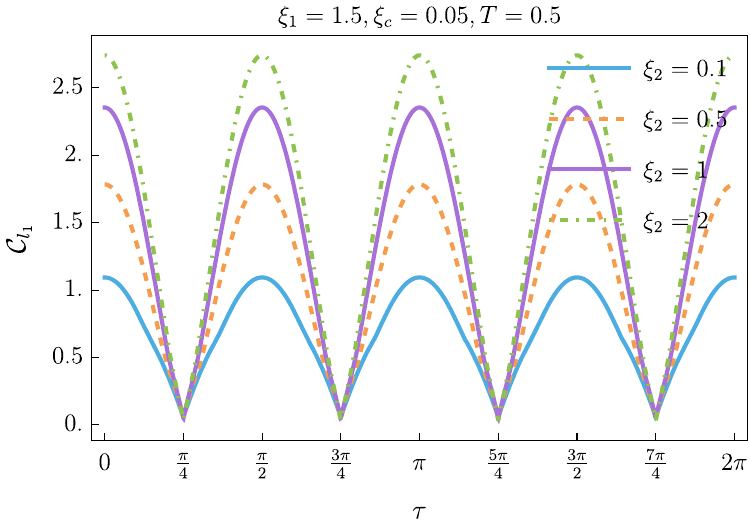}}
			\caption{Variation of (a) ergotropy $\mathcal{E}$, (b) instantaneous power $\mathcal{P}$, (c) capacity $\mathcal{K}$, and (d) $l_1$-norm of coherence $\mathcal{C}_{l_1}$ as a function of $ \tau=\Omega t$ for different values of $\xi_2$. For all plots, the fixed parameters are $\xi_1 =1.5$, $\xi_c=0.05$, and $T=0.5$.}
			\label{figure1}
   \end{figure*}

		\begin{figure*}[t]
			\centering
			\subfigure[]{\label{figure2a}\includegraphics[scale=0.5]{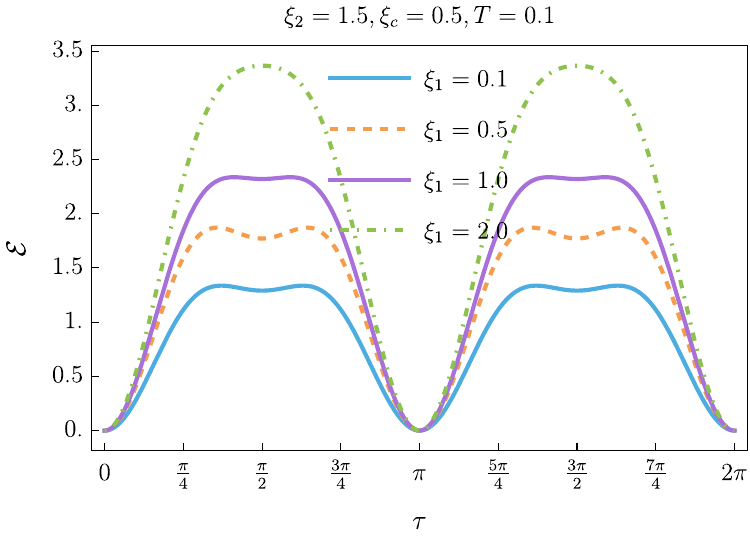}}
			\subfigure[]{\label{figure2b}\includegraphics[scale=0.5]{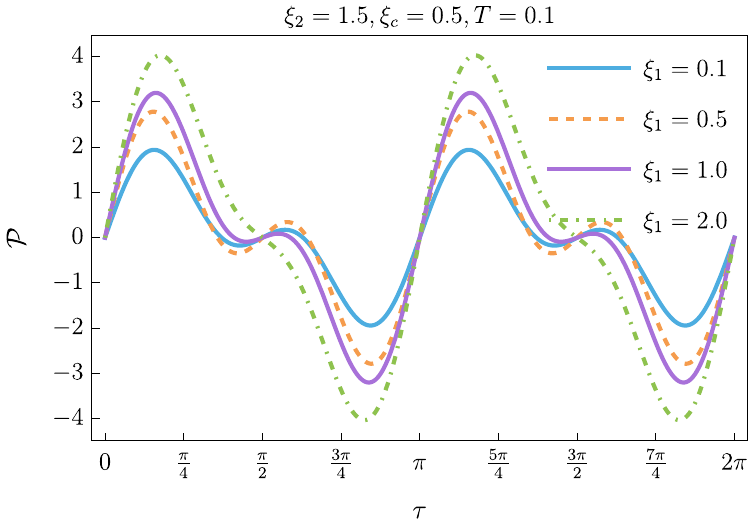}}
   \subfigure[]{\label{figure2c}\includegraphics[scale=0.5]{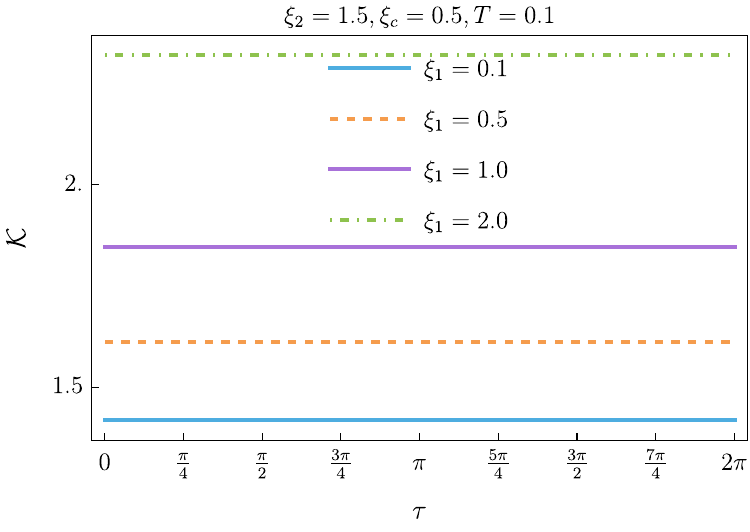}}
			\subfigure[]{\label{figure2d}\includegraphics[scale=0.5]{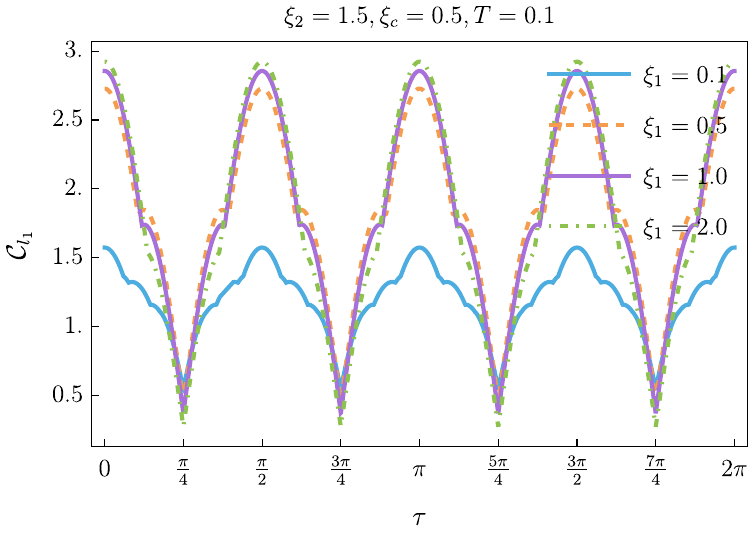}}
			\caption{Variation of (a) ergotropy $\mathcal{E}$, (b) instantaneous power $\mathcal{P}$, (c) capacity $\mathcal{K}$, and (d) $l_1$-norm of coherence $\mathcal{C}_{l_1}$ as a function of $ \tau=\Omega t$ for different values of $\xi_1$. For all plots, the fixed parameters are $\xi_2 =1.5$, $ \xi_c=0.5$, and $T=0.1$.}
			\label{figure2}
		\end{figure*}
\section{Results and discussion \label{sec2}}
In this section, we investigate the charging performance of the SQB under various physical parameter regimes. We compute and plot the ergotropy $\mathcal{E}$, instantaneous power $\mathcal{P}$, capacity $\mathcal{K}$, and the $l_1$-norm of coherence $\mathcal{C}_{l_1}$. First, we vary the Josephson energies $\xi_{1,2}$ between the qubits, and then we examine the battery performance under different strengths of the mutual coupling energy $\xi_c$. Finally, we explore the significance of the mutual coupling energy $\xi_c$ in the system and its role in the efficient charging of the SQB. We consider both equal and varying Josephson energies in the SQB model.

In Fig. \ref{figure1}, we present the variations of key quantities associated with our model: ergotropy $\mathcal{E}$, instantaneous power $\mathcal{P}$, capacity $\mathcal{K}$, and the $l_1$-norm of coherence $\mathcal{C}_{l_1}$ as functions of $\tau = \Omega t$ for different values of $\xi_2$. The parameter $\xi_2$ represents the Josephson energy of one of the qubits, which influences its individual energy scale and contributes to the overall charging dynamics of the quantum battery. The fixed parameter $\xi_1 = 1.5$ corresponds to the Josephson energy of the second qubit, playing a similar role in defining the system’s energy distribution. The mutual coupling energy $\xi_c = 0.05$ determines the strength of interaction between the two qubits, impacting both energy transfer efficiency and coherence properties. The temperature parameter $T = 0.5$ governs thermal fluctuations, where higher values lead to increased decoherence effects.

To systematically analyze the impact of $\xi_2$, we consider four representative values: $\xi_2 = 0.1, 0.5, 1$, and $2$. These choices span a range from weak ($\xi_2 = 0.1$) to strong ($\xi_2 = 2$) Josephson energy, allowing us to explore different operating regimes of the quantum battery. At lower values of $\xi_2$, the energy gap of the qubit is smaller, leading to slower charging dynamics, while at higher values, the system enters a regime of enhanced energy storage and coherence effects. This selection enables us to contrast scenarios where the qubit behaves as a weakly or strongly coupled element within the battery system. A more detailed analysis of these parameter choices is provided in Appendix~\ref{appD}.

Figure \ref{figure1}(a) depicts the plots of ergotropy, $\mathcal{E}$, as a function of $\tau$. Ergotropy, representing the extractable work from the quantum system, displays oscillatory behavior, signifying energy fluctuations within the system's quantum state. The amplitude of these oscillations increases with $\xi_2$, the Josephson energy of the second qubit. This suggests that larger values of $\xi_2$ enhance the system's ability to store and extract work, making it more efficient for quantum energy storage. These results imply that tuning the Josephson energy of the second qubit could optimize the quantum battery's performance in terms of maximizing its stored work.

Figure \ref{figure1}(b) illustrates the instantaneous power, $\mathcal{P}$, which reflects the rate at which energy is either being extracted from or injected into the system. Positive values of $\mathcal{P}$ correspond to charging the quantum battery, while negative values indicate discharging. The oscillatory nature of $\mathcal{P}$ corresponds to the behavior of ergotropy in Fig. \ref{figure1}(a). For instance, during the interval $\tau \in (0; \pi/2)$, $\mathcal{P} > 0$, meaning the system is charging the quantum battery. This aligns with the increasing ergotropy over the same interval, which achieves the maximum value for $\tau = \pi/2$. Conversely, for $\tau\in(\pi/2; \pi)$, $\mathcal{P}$ becomes negative, indicating that the system is discharging, as reflected by the decreasing ergotropy. At $\tau = \pi$, ergotropy drops to zero, marking the point where the system has fully discharged, and the sign of $\mathcal{P}$ flips, indicating the start of a new charging cycle. This periodic behavior of charging and discharging continues, governed by the system’s oscillatory dynamics. The dependence of power, $\mathcal{P}$, on $\xi_2$ is significant; larger values of $\xi_2$ lead to a noticeable increase in peak oscillation values. This behavior demonstrates how system parameters enhance battery charging performance, resulting in a rapid transfer of energy to the battery.

Fig. \ref{figure1}(c) displays the capacity $\mathcal{K}$ as a function of $\tau$. In contrast to the other quantities, the capacity remains a constant function in the domain of $\tau$. However, the constant value of $\mathcal{K}$ increases with larger $\xi_2$, indicating that enhanced Josephson energy is more effective in distributing energy uniformly across the battery, leading to a stable increase in the extractable energy. Although most properties of battery capacity can be derived from ergotropy, we argue that battery capacity is a more fundamental quantity, as it remains unchanged during unitary charging or discharging.

Fig. \ref{figure1}(d) examines the $l_1$-norm of coherence $\mathcal{C}_{l_1}$. The coherence exhibits a more complex behavior, with clear periodic features reflecting the underlying quantum interference effects. As $\xi_2$ increases, the coherence shows enhanced oscillatory behavior, suggesting that increased coupling may facilitate better preservation of quantum coherence in the system. This coherence is essential for maximizing the performance of quantum protocols, further illustrating the importance of parameter selection in the optimization of quantum resources.

Overall, Fig. \ref{figure1} shows the interplay between ergotropy, instantaneous power, capacity, and coherence. The results indicate that careful tuning of the parameters, particularly $\xi_2$, can significantly influence both the efficiency of energy extraction and the preservation of quantum coherence, which are critical for achieving high efficiency in quantum batteries.

In Fig. \ref{figure2}, we present the variations of ergotropy $\mathcal{E}$, instantaneous power $\mathcal{P}$, capacity $\mathcal{K}$, and the $l_1$-norm of coherence $\mathcal{C}_{l_1}$ as functions of $\tau = \Omega t$ for varying values of $\xi_1$. The other parameters are fixed with $\xi_2 = 1.5$, $\xi_c = 0.5$, and $T = 0.1$.

Fig. \ref{figure2}(a) presents the ergotropy $\mathcal{E}$ as a function of $\tau$. While the general oscillatory behavior is similar to that observed in Fig. \ref{figure1}(a), the patterns differ due to the distinct roles played by $\xi_1$ and $\xi_2$. In Fig. \ref{figure2}(a), $\xi_1$, the Josephson energy of the first qubit, is varied while $\xi_2$ is held fixed. As $\xi_1$ increases, the amplitude of ergotropy oscillations grows, indicating an enhancement in the SQB's work extraction potential. This result emphasizes the significance of $\xi_1$ in determining the system's energy dynamics, as it directly governs the properties of the first qubit. However, the influence of $\xi_1$ is modulated by the coupling energy $\xi_c$, which governs the interaction strength between the two qubits. The higher value of $\xi_c$ in Fig. \ref{figure2}(a) ($\xi_c = 0.5$) results in more diverse shapes, reflecting the stronger interplay between the qubits. In contrast, Fig. \ref{figure1}(a) shows smoother oscillations due to the weaker coupling ($\xi_c = 0.05$).

Fig. \ref{figure2}(b) illustrates the instantaneous power $\mathcal{P}$ as a function of $\tau$, showing a similar oscillatory behavior as in Fig. \ref{figure1}(b), but with a notable increase in complexity. The variation of $\xi_1$ affects the power dynamics differently than $\xi_2$ due to the stronger coupling strength $\xi_c$. Here, the non-sinusoidal oscillations become more noticeable, indicating enhanced energy exchange between the two qubits. The peak values of instantaneous power also grow with increasing $\xi_1$, reinforcing the role of this parameter in optimizing energy extraction. The higher coupling strength $\xi_c$ amplifies the interaction between the qubits, causing their dynamics to be more interdependent and resulting in a more complex power profile compared to Fig. \ref{figure1}(b).

The comparison between Figs. \ref{figure1} and \ref{figure2} in terms of the ergotropy and the instantaneous power highlights the interplay between the Josephson energies and the coupling strength. In Fig. \ref{figure1}, where $\xi_c$ is weaker, the influence of $\xi_2$ dominates, leading to smoother oscillatory behaviors. In contrast, the stronger coupling in Fig. \ref{figure2} redistributes the influence of $\xi_1$ and $\xi_2$, resulting in more complex dynamics. These observations demonstrate that $\xi_c$ plays a crucial role in shaping the energy and power dynamics of the system and it must be carefully considered in the design of efficient quantum batteries.

		\begin{figure*}[t]
			\centering
			\subfigure[]{\label{figure3a}\includegraphics[scale=0.5]{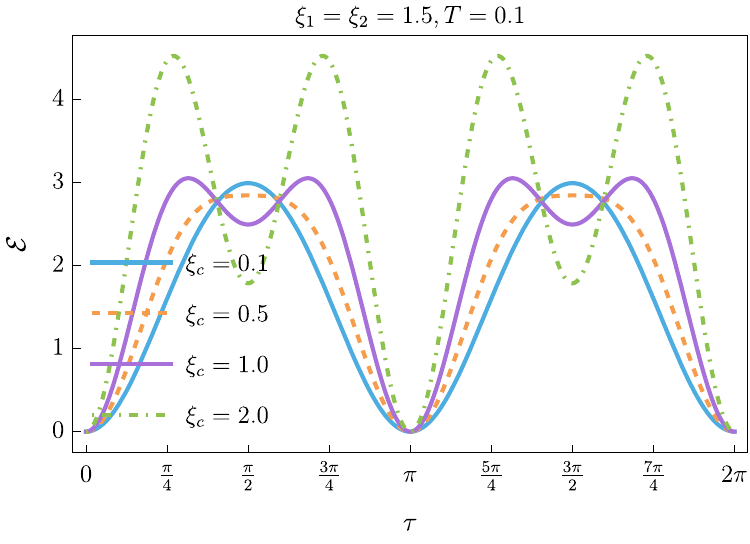}}
			\subfigure[]{\label{figure3b}\includegraphics[scale=0.5]{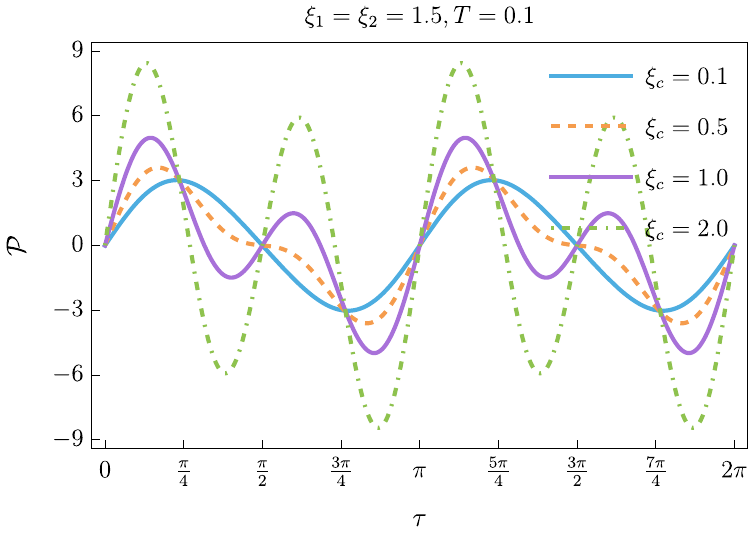}}
            \subfigure[]{\label{figure3c}\includegraphics[scale=0.5]{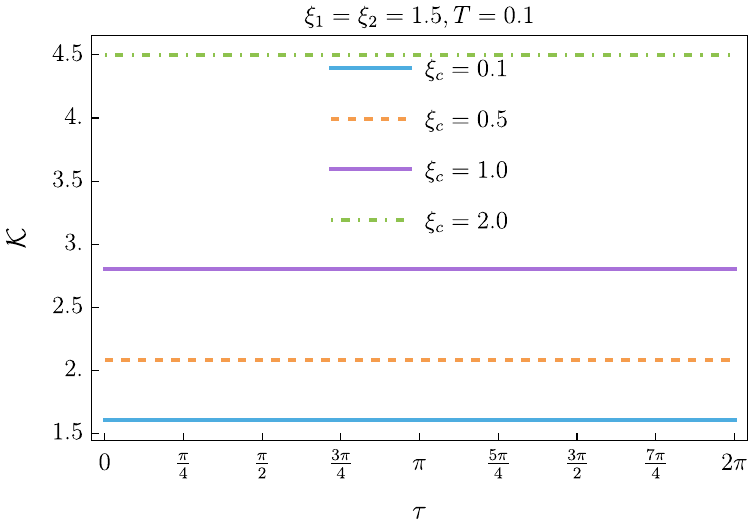}}
			\subfigure[]{\label{figure3d}\includegraphics[scale=0.5]{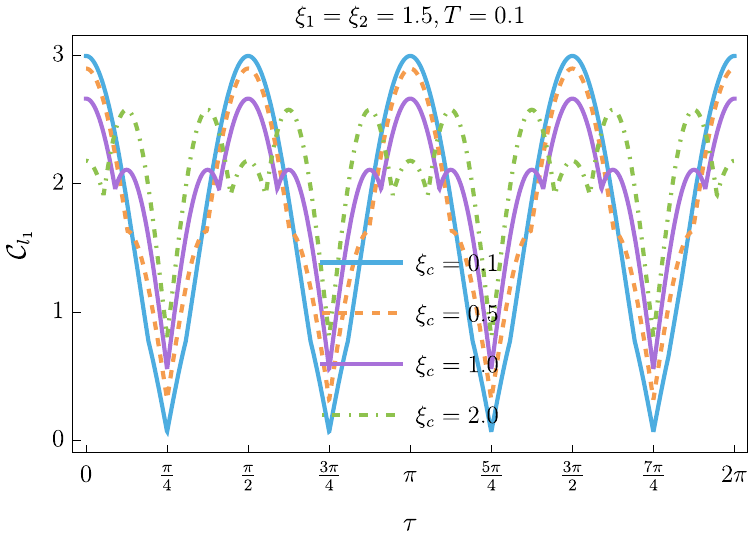}}
			\caption{Variation of (a) ergotropy $\mathcal{E}$, (b) instantaneous power $\mathcal{P}$, (c) capacity $\mathcal{K}$, and (d) $l_1$-norm of coherence $\mathcal{C}_{l_1}$ as a function of $ \tau=\Omega t$ for different values of $\xi_c$. For all subfigures, $\xi_1=\xi_2 =1.5$ and $T=0.1$.}
			\label{figure3}
		\end{figure*}

Fig. \ref{figure2}(c) shows the capacity $\mathcal{K}$ as a function of $\tau$. As in Fig. \ref{figure1}(c), capacity is a constant function in the range of $\tau$, with higher $\xi_1$ generally leading to a greater increase in capacity in the absence of transient oscillations. This stability of capacity, despite fluctuations in ergotropy and power, indicates that an SQB can store and deliver energy in different parameter regimes, as shown by the observations in Fig. \ref{figure1}. However, in our SQB charging model, the system’s capacity is mainly decided by the Josephson energy of the qubit.

Fig. \ref{figure2}(d) examines the $l_1$-norm of coherence $\mathcal{C}_{l_1}$. The coherence exhibits oscillatory behavior, paralleling the dynamics observed in Fig. \ref{figure1}(d). As $\xi_1$ increases, the oscillations become more pronounced, indicating that higher values of $\xi_1$ enhance the coherence properties of the quantum state. However, this improvement appears to plateau at higher values of $\xi_1$, as the corresponding plots converge and nearly overlap, suggesting a saturation effect where further increases in $\xi_1$ do not lead to additional coherence enhancement.

The results in Fig. \ref{figure2} reflect trends observed in Fig. \ref{figure1}, highlighting the importance of parameter selection for controlling the charging time of the SQB to ensure maximum charging energy. These parameters influence ergotropy, instantaneous power, and coherence oscillator frequency, which are critical for maximizing energy transfer and achieving higher transient capacity values. The interplay among ergotropy, instantaneous power, capacity, and coherence demonstrates that the SQB can store energy more quickly and has a greater maximum extractable work, as quantified by ergotropy.

In Fig. \ref{figure3}, we explore the variations of ergotropy $\mathcal{E}$, instantaneous power $\mathcal{P}$, capacity $\mathcal{K}$, and the $l_1$-norm of coherence $\mathcal{C}_{l_1}$ as functions of $\tau = \Omega t$ for varying values of $\xi_c$. Specifically, we consider $\xi_c = 0.1, \,0.5,\, 1$, and $2$, which allows us to analyze a range of coupling strengths from weak to strong interactions. Lower values of $\xi_c$ correspond to minimal energy exchange between qubits, while higher values enhance interaction effects, influencing both coherence properties and energy transfer efficiency. The remaining parameters are fixed with $\xi_1 = \xi_2 = 1.5$ and $T = 0.1$.

		\begin{figure*}[t]
			\centering
			\subfigure[]{\label{figure1a}\includegraphics[scale=0.5]{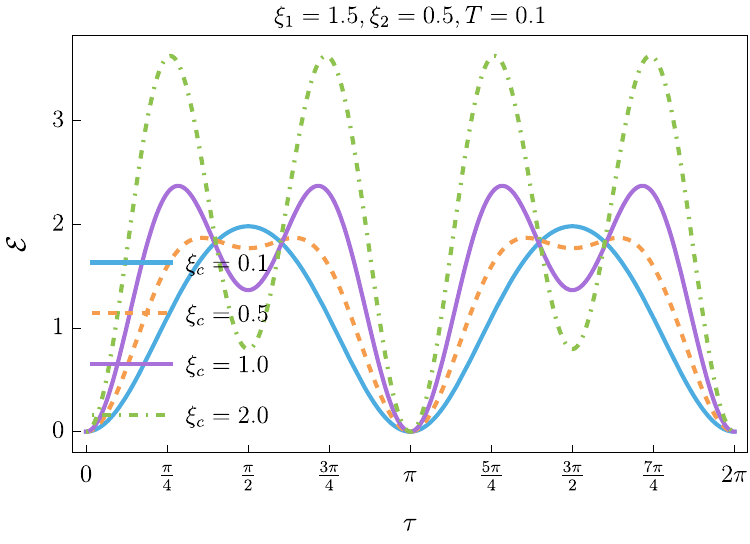}}
			\subfigure[]{\label{figure1b}\includegraphics[scale=0.5]{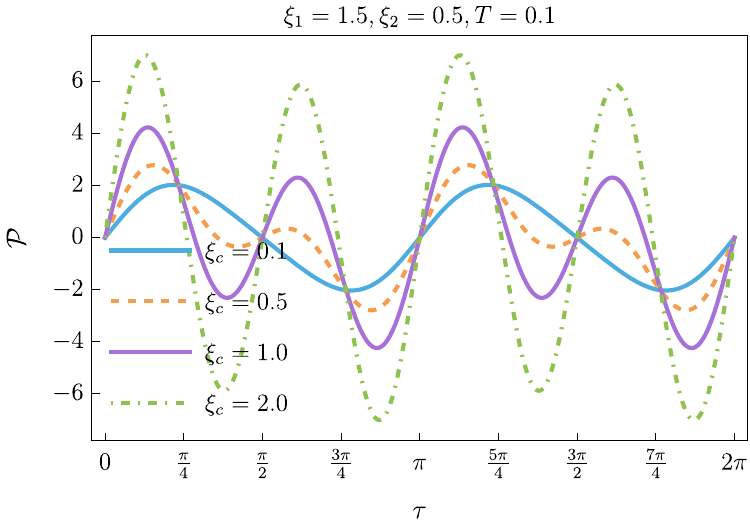}}
            \subfigure[]{\label{figure1c}\includegraphics[scale=0.5]{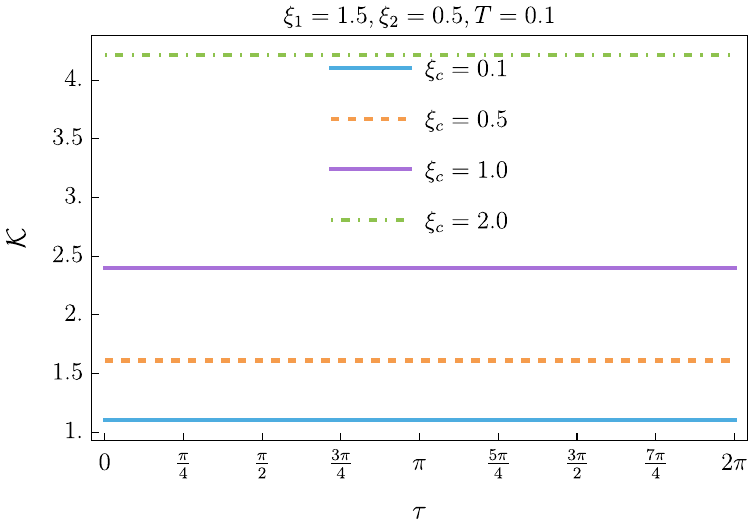}}
			\subfigure[]{\label{figure1d}\includegraphics[scale=0.5]{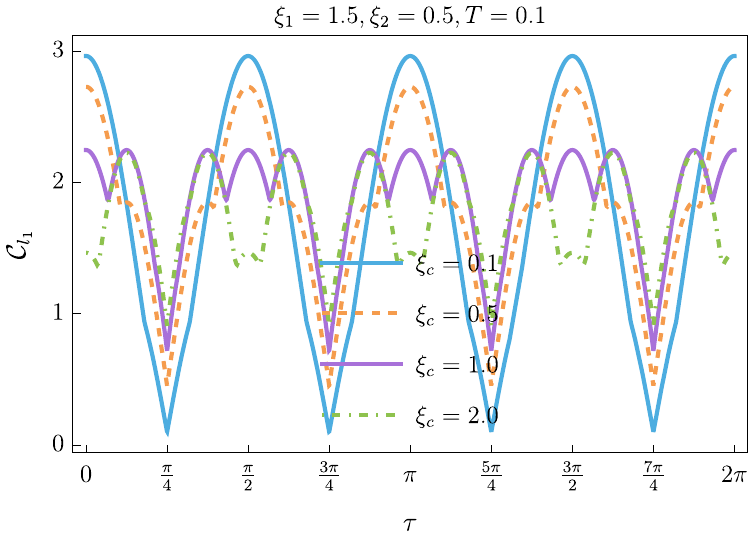}}
			\caption{Variation of (a) ergotropy $\mathcal{E}$, (b) instantaneous power $\mathcal{P}$, (c) capacity $\mathcal{K}$ and (d) $l_1$-norm of coherence $\mathcal{C}_{l_1}$ as a function of $ \tau=\Omega t$ for different values of $\xi_c$. For all subfigures, $\xi_1=1.5$, $\xi_2 =0.5$, and $T=0.1$.}
			\label{figure4}
		\end{figure*}

Fig. \ref{figure3}(a) depicts the plots of ergotropy $\mathcal{E}$ with respect to $\tau$. As observed in Figs. \ref{figure1}(a) and \ref{figure2}(a), the ergotropy exhibits oscillatory behavior. Notably, for higher values of $\xi_c$, not only does the amplitude of oscillation increase, but the frequency of oscillation also rises. This phenomenon indicates that larger values of the coupling energy between the qubits allow for more frequent cycles of charging and discharging within the same time interval. Physically, this can be interpreted as the system becoming more dynamic, where the energy states transition more rapidly, thereby enhancing the potential for more rapid charging of the quantum battery.

Fig. \ref{figure3}(b) shows the instantaneous power $\mathcal{P}$, which continues to exhibit similar oscillatory behavior as observed in Figs. \ref{figure1}(b) and \ref{figure2}(b). The increased amplitude and frequency of oscillation with larger $\xi_c$ parallel the trends noted in ergotropy. This suggests that the power output of the system also improves, providing more energy within a given time frame. The correlation between ergotropy and instantaneous power reinforces the observation that an enhanced coupling energy directly translates into a greater efficiency of the SQB.

Fig. \ref{figure3}(c) illustrates the capacity $\mathcal{K}$, which remains constant across the various values of $\tau$, as in Figs. \ref{figure1}(c) and \ref{figure2}(c). However, the value of $\mathcal{K}$ is directly influenced by $\xi_c$: increasing $\xi_c$ leads to a higher constant value of capacity throughout the domain of $\tau$. This suggests that while energy extraction fluctuates with ergotropy and power, optimal performance is achieved with larger $\xi_c$, ensuring stable and robust performance across the range of $\tau$.

Fig. \ref{figure3}(d) examines the $l_1$-norm of coherence $\mathcal{C}_{l_1}$. Similar to the behavior observed in Figs. \ref{figure1}(d) and \ref{figure2}(d), $\mathcal{C}_{l_1}$ shows oscillatory behavior as $\xi_c$ increases. However, unlike ergotropy and instantaneous power, the amplitude of the oscillations in coherence slightly declines with increasing $\xi_c$, despite the frequency of oscillation increasing. This decline in amplitude suggests that although the qubits become more strongly coupled, leading to more frequent state transitions (i.e., more maxima), the overall ability of the system to maintain a high level of coherence diminishes slightly. Physically, this could reflect the increased interaction between the qubits causing more rapid fluctuations in coherence but limiting the maximum achievable coherence at any given time.

In general, the results presented in Fig.~\ref{figure3} highlight the impact of the coupling energy $\xi_c$ between qubits on the system's energy dynamics and coherence properties. In particular, increasing $\xi_c$ enhances the frequency of oscillations in ergotropy, instantaneous power, and coherence, indicating that the system becomes more dynamic and responsive. However, while both ergotropy and power exhibit increases in both amplitude and frequency, coherence shows an interesting trade-off: the frequency of oscillations increases, but the amplitude slightly declines. This suggests that stronger qubit coupling facilitates more rapid state transitions, but at the cost of slightly reducing the overall coherence levels. This observation showcases a delicate balance between coherence and energy extraction in strongly coupled quantum systems.

Finally, in Fig. \ref{figure4}, we analyze the variations of ergotropy $\mathcal{E}$, instantaneous power $\mathcal{P}$, capacity $\mathcal{K}$, and the $l_1$-norm of coherence $\mathcal{C}_{l_1}$ as functions of $\tau = \Omega t$ for different values of the coupling strength $\xi_c$. The other parameters are fixed with $\xi_1 = 1.5$, $\xi_2 = 0.5$, and $T = 0.1$.

Fig. \ref{figure4}(a) depicts the plots of ergotropy $\mathcal{E}$ with respect to $\tau$. While the overall tendencies are similar to those in Fig. \ref{figure3}(a), we observe that the maximum values of ergotropy are significantly lower, which means that the maximum amount of energy stored in the SQB is reduced. This is due to the smaller value of $\xi_2$, which represents the Josephson energy of the second qubit. With $\xi_1 = 1.5$ and $\xi_2 = 0.5$, the asymmetry in the Josephson energies of the qubits reduces the system’s ability to extract work. The oscillatory nature persists, with both the amplitude and frequency increasing as $\xi_c$ increases. However, the reduction in $\xi_2$ limits the overall ergotropy, as the second qubit contributes less energy to the system.

Fig. \ref{figure4}(b) shows the instantaneous power $\mathcal{P}$. Similar to the trends observed in ergotropy, the maximum values of $\mathcal{P}$ are lower compared to Fig. \ref{figure3}(b). This again reflects the reduced Josephson energy of the second qubit, which lowers the power output of the system. The oscillations in power increase in both amplitude and frequency as $\xi_c$ increases, consistent with the behavior of ergotropy. The coupling strength $\xi_c$ enhances the rate of energy transfer between the qubits, but the total available energy is constrained by the smaller value of $\xi_2$.

Fig. \ref{figure4}(c) illustrates the capacity $\mathcal{K}$, which also shows lower overall values. The SQB capacity is hardly influenced by the reduced Josephson energy of the second qubit, $\xi_2 = 0.5$, which lowers the system's ability to store and transfer energy. In addition, the capacity is more sensitive to coupling strength $\xi_c$ in Fig. \ref{figure3}(c), the increase of $\xi_c$ will increase the SQB capacity. It means that we could enhance the capacity by increasing the coupling strength and the Josephson energies of the qubits.

Fig. \ref{figure4}(d) examines the $l_1$-norm of coherence, denoted as $\mathcal{C}_{l_1}$. The general behavior of coherence is similar to that in Fig. \ref{figure3}(d), with oscillations increasing in frequency as $\xi_c$ rises. Additionally, the amplitude of the oscillations in $\mathcal{C}_{l_1}$ declines with increasing values of $\xi_c$. While the coupling strength enhances the frequency of coherence oscillations, the asymmetry between the qubits’ Josephson energies ($\xi_1 = 1.5$ and $\xi_2 = 0.5$) leads to a reduction in the maximum achievable coherence compared to Fig. \ref{figure3}(d). However, there is also a clear difference to be noticed. The symmetrical Josephson energies yield significantly higher coherence.

The results in Fig. \ref{figure4} highlight the impact of asymmetry in the Josephson energies of the two qubits on the system's performance. With $\xi_1 = 1.5$ and $\xi_2 = 0.5$, the overall ergotropy, instantaneous power, and capacity are significantly reduced compared to the results in Fig. \ref{figure3}. This observation confirms that the energy imbalance between the qubits limits their collective ability to extract work, generate power, and store information, thereby influencing the overall performance of the SQB. The coupling strength parameter $\xi_c$ is beneficial as it increases the frequency of oscillations in ergotropy, power, and coherence, demonstrating that stronger coupling leads to more dynamic interactions between the qubits.

\section{Scalability of SQBs \label{sec3}}
Scalability is a crucial aspect of practical quantum battery implementations, particularly in superconducting circuit architectures. While our analysis focuses on SQB as a fundamental building block, the principles established in this work can be extended to multi-qubit configurations.

One promising method for scalability is the utilization of collective charging protocols in multi-qubit superconducting systems. Previous studies have shown that collective effects, such as quantum synchronization and cooperative energy transfer, can lead to enhanced charging power and improved energy storage efficiency in quantum batteries \cite{ferraro2018high,PhysRevLett.122.047702}. In a superconducting circuit, such collective charging could be realized through coherent coupling between multiple transmon qubits or by employing superconducting resonators as mediators for global interactions \cite{PhysRevApplied.14.024092}.

Another approach involves leveraging modular circuit architectures that integrate superconducting qubit arrays into larger networked configurations. Recent advances in superconducting quantum computing have demonstrated the feasibility of scaling up qubit arrays with high coherence times and tunable interactions, which could be exploited for enhanced quantum battery performance \cite{Krantz2019,Arute2019}. By optimizing the connectivity between qubits and tuning Josephson energies appropriately, it may be possible to sustain coherent energy transfer across a larger quantum battery system while mitigating decoherence effects.

Additionally, flux-tunable qubits and superconducting metamaterials provide further scalability options by enabling dynamic control over interactions, allowing for adaptive charging protocols that optimize energy storage efficiency as the system scales \cite{kjaergaard2020superconducting,Devoret2013}.

Future investigations should focus on the robustness of these multi-qubit architectures under realistic noise conditions and their potential advantages over classical and semi-classical charging mechanisms. Understanding the interplay between coherence, entanglement, and charging efficiency in large-scale superconducting quantum batteries will be essential for advancing their practical implementation.

\section{Experimental feasibility and optimization strategies \label{sec4}}
Our study is grounded in realistic superconducting circuit parameters, ensuring that the proposed SQB model aligns with current experimental capabilities in superconducting qubit technology. Recent advancements in superconducting quantum circuits have demonstrated precise tunability of coupling strengths and energy levels, enabling controlled interactions between qubits \cite{Krantz2019,Devoret2013}. Moreover, the analysis considers the tunability of Josephson energies and coupling strengths, both of which can be dynamically controlled in state-of-the-art superconducting platforms using external magnetic flux and circuit design optimizations \cite{Nori2011,Wendin2017}. This flexibility ensures that our model can be implemented using existing fabrication techniques. Interestingly, the oscillatory behavior of ergotropy and power observed in our model is directly linked to experimentally accessible coherence times and coupling strengths in superconducting qubits. These characteristics have been extensively studied in transmon-based superconducting circuits, which exhibit coherence times on the order of tens to hundreds of microseconds, allowing sufficient time for energy transfer and extraction \cite{PhysRevA.76.042319,kjaergaard2020superconducting}.

Regarding the optimization strategies, we demonstrate that tuning the coupling strength $\xi_c$ enhances the frequency and amplitude of energy oscillations, offering a method to optimize power output in practical applications. This is consistent with previous studies showing that strong coupling enhances quantum energy transfer efficiency in superconducting systems \cite{PRXQuantum.2.010339}. Our results indicate that an optimal balance between the Josephson energies $\xi_1$ and $\xi_2$ is crucial for maximizing energy storage and extraction. Imbalances in Josephson energies create asymmetries in the system’s energy levels, which can limit the maximum extractable work. Similar effects have been observed in other superconducting qubit-based quantum thermodynamic models \cite{Pekola2015}. By analyzing the interplay between coherence, energy transfer, and information capacity, we offer insights into designing SQBs with improved performance, particularly in mitigating losses due to parameter asymmetry. Our findings align with prior research on coherence-driven quantum energy storage, reinforcing the role of quantum coherence as a resource for energy efficiency \cite{alicki2013entanglement,binder2015quantacell}.

Thus, this paper provides a theoretical foundation for engineering SQBs with controlled energy dynamics, paving the way for experimental validation and further optimization in quantum energy storage systems. With continued advancements in superconducting qubit fabrication and control, experimental demonstrations of SQBs based on coupled qubits may soon be realized, delivering optimal performance \cite{ferraro2018high,hu2022optimal,PhysRevLett.118.150601}.

\section{Discussion and Conclusions \label{sec5}}
In this paper, we investigated the performance of SQB charging using a quantum system composed of two superconducting coupled qubits, specifically focusing on the variation of ergotropy $\mathcal{E}$, instantaneous power $\mathcal{P}$, capacity $\mathcal{K}$, and the $l_1$-norm of coherence $\mathcal{C}_{l_1}$. By systematically varying the coupling strength $\xi_c$ as well as the Josephson energies of the qubits $\xi_1$ and $\xi_2$, we have explored how these parameters influence the performance of the quantum battery.

Our analysis demonstrates several significant findings. First, we observe that for all values of the coupling strength $\xi_c$, the system exhibits oscillatory behavior in ergotropy, instantaneous power, and coherence. These oscillations reflect the underlying quantum energy transfer dynamics between the two qubits. As $\xi_c$ increases, both the amplitude and frequency of these oscillations increase. This tendency means that stronger coupling enhances the energy exchange between the qubits and results in more rapid transitions within the system. The large coupling strength $\xi_c$ corresponds to stronger coherent interactions between the two qubits, resulting in higher-frequency oscillations in ergotropy, power, and coherence during the charging process. Efficient energy transfer can be adjusted to a significant value by selecting the coupling strength $\xi_c$ appropriately. However, in cases where there is an asymmetry between the Josephson energies of the qubits, particularly when $\xi_2$ is significantly smaller than $\xi_1$, the maximum achievable ergotropy and power are limited.  This is because a significant disparity in $\xi_1$ and $\xi_2$ affects the energy gaps of the system, limiting the extent to which energy can be cyclically stored and extracted. The observed reduction in peak ergotropy and power is thus not a suppression of energy exchange per se but rather a constraint on how effectively energy can be harnessed for work extraction due to the system’s altered energy landscape. This implies that the Josephson energies of the qubits must be equal to ensure maximum energy transfer.

In terms of information capacity $\mathcal{K}$, we find that this quantity varies with each value of $\xi_c$ and is influenced by the Josephson energies of the qubits. This variation reflects the fact that the performance of the system is closely tied to both the energy properties of the qubits and the coupling strength. Systems with higher Josephson energies exhibit greater capacity, enabling them to store more energy. However, the results also indicate that when one qubit has significantly lower Josephson energy than the other (as shown in Fig. \ref{figure4}), the system’s overall information capacity is reduced.

Thus, to achieve optimal objective functions, a detailed analysis of SQB parameters is considered valuable.


\appendix

\section{Elements of thermal state}
\label{AppA}
The elements of thermal state \eqref{eq:5} are

\begin{equation}
\mathcal{R}_{th_{11}}=\mathcal{R}_{th_{44}} =\frac{A_-+A_+-\frac{2 B_- \xi _c}{\alpha _-}-\frac{2 B_+ \xi _c}{\alpha _+}}{4 \left(A_-+A_+\right)},
\end{equation}

\begin{equation}
    \mathcal{R}_{th_{12}} =\mathcal{R}_{th_{21}} =
\frac{\frac{B_- \left(\xi _2-\xi _1\right)}{2 \alpha _-}+\frac{B_+ \left(\xi _1+\xi _2\right)}{2 \alpha _+}}{2 \left(A_-+A_+\right)},
\end{equation}

\begin{equation}
    \mathcal{R}_{th_{13}} =\mathcal{R}_{th_{31}} =\mathcal{R}_{th_{24}} =\mathcal{R}_{th_{42}} =\frac{\frac{B_- \left(\xi _1-\xi _2\right)}{2 \alpha _-}+\frac{B_+ \left(\xi _1+\xi _2\right)}{2 \alpha _+}}{2 \left(A_-+A_+\right)},
\end{equation}

\begin{equation}
    \mathcal{R}_{th_{14}} =\mathcal{R}_{th_{41}} = \frac{-A_-+A_++\frac{2 B_- \xi _c}{\alpha _-}-\frac{2 B_+ \xi _c}{\alpha _+}}{4 \left(A_-+A_+\right)},
\end{equation}

\begin{equation}
    \mathcal{R}_{th_{22}}=\mathcal{R}_{th_{33}} = \frac{A_-+A_++2 \left(\frac{B_-}{\alpha _-}+\frac{B_+}{\alpha _+}\right) \xi _c}{4 \left(A_-+A_+\right)},
\end{equation}

\begin{equation}
    \mathcal{R}_{th_{23}} = \mathcal{R}_{th_{32}} = \frac{-A_-+A_+-\frac{2 B_- \xi _c}{\alpha _-}+\frac{2 B_+ \xi _c}{\alpha _+}}{4 \left(A_-+A_+\right)},
\end{equation}

\begin{equation}
    \mathcal{R}_{th_{34}} = \mathcal{R}_{th_{43}} = \frac{\frac{B_- \left(\xi _2-\xi _1\right)}{2 \alpha _-}+\frac{B_+ \left(\xi _1+\xi _2\right)}{2 \alpha _+}}{2 \left(A_-+A_+\right)},
\end{equation}
where the partition function is given by
\begin{equation}
    \mathcal{Z}=2 \left(A_{+}+A_{-}\right),
\end{equation}
with $\alpha_\pm = \sqrt{4 \xi_{c}^2+(\xi_{1}\pm\xi_{2})^2},$ $A_{\pm}=\cosh \left(\frac{\alpha_\pm}{2 T}\right),$ and
$B_{\pm}=\sinh \left(\frac{\alpha_\pm}{2 T}\right).$

\begin{widetext}
\section{Density matrix
$\mathcal{R_X}(t)$ \label{appB}}

The density matrix
$\mathcal{R_X}(t)$ \eqref{RX} is given by
\begin{equation}
\mathcal{R_X}(t) = \begin{pmatrix}
\mathcal{R_X}_{11}(t)  & \mathcal{R_X}_{12}(t)  & \mathcal{R_X}_{13}(t)  & \mathcal{R_X}_{14}(t)  \\
\mathcal{R_X}_{21}(t)  & \mathcal{R_X}_{22}(t)  & \mathcal{R_X}_{23}(t)  & \mathcal{R_X}_{24}(t)  \\
\mathcal{R_X}_{31}(t)  & \mathcal{R_X}_{32}(t)  & \mathcal{R_X}_{33}(t)  & \mathcal{R_X}_{34}(t)  \\
\mathcal{R_X}_{41}(t)  & \mathcal{R_X}_{42}(t)  & \mathcal{R_X}_{43}(t)  & \mathcal{R_X}_{44}(t)
\end{pmatrix},
\end{equation}
where

\small
\begin{equation}
   \mathcal{R_X}_{11}(t)=\mathcal{R_X}_{44}(t) = -\frac{-2 A_- \cos ^2(2 \tau )+A_+ (\cos (4 \tau )-3)+\frac{4 B_- \xi _c \cos ^2(2 \tau )}{\alpha _-}+\frac{2 B_+ \left(\xi _c \cos (4 \tau
   )+\xi _c-2 \left(\xi _1+\xi _2\right) \sin (2 \tau )\right)}{\alpha _+}}{8 \left(A_-+A_+\right)},
\end{equation}

\small
\begin{equation}
    \mathcal{R_X}_{12}(t) = \mathcal{R_X}_{21}(t) = \frac{\cos (2 \tau ) \left(\sin (2 \tau ) \left(A_--A_+-\frac{2 B_- \xi _c}{\alpha _-}-\frac{2 B_+ \xi _c}{\alpha _+}\right)+\frac{B_-
   \left(\xi _2-\xi _1\right)}{\alpha _-}+\frac{B_+ \left(\xi _1+\xi _2\right)}{\alpha _+}\right)}{4 \left(A_-+A_+\right)},
\end{equation}

\small
\begin{equation}
    \mathcal{R_X}_{13}(t) =\mathcal{R_X}_{31}(t) = \frac{\cos (2 \tau ) \left(\sin (2 \tau ) \left(A_--A_+-\frac{2 B_- \xi _c}{\alpha _-}-\frac{2 B_+ \xi _c}{\alpha _+}\right)+\frac{B_-
   \left(\xi _1-\xi _2\right)}{\alpha _-}+\frac{B_+ \left(\xi _1+\xi _2\right)}{\alpha _+}\right)}{4 \left(A_-+A_+\right)},
\end{equation}

\small
\begin{equation}
   \mathcal{R_X}_{14}(t) = \mathcal{R_X}_{41}(t) = \frac{-2 A_- \cos ^2(2 \tau )+2 A_+ \cos ^2(2 \tau )+\frac{4 B_- \xi _c \cos ^2(2 \tau )}{\alpha _-}+\frac{2 B_+ \xi _c (\cos (4 \tau
   )-3)}{\alpha _+}}{8 \left(A_-+A_+\right)},
\end{equation}

\small
\begin{equation}
    \mathcal{R_X}_{22}(t) = \frac{\alpha _+ \left(-\left(\alpha _- A_- (\cos (4 \tau )-3)\right)+\alpha _- A_+ (\cos (4 \tau )+1)+4 B_- \left(\xi _2-\xi _1\right) \sin (2
   \tau )\right)+4 \left(\alpha _- B_++\alpha _+ B_-\right) \xi _c \cos ^2(2 \tau )}{8 \alpha _- \alpha _+ \left(A_-+A_+\right)},
\end{equation}

\small
\begin{equation}
    \mathcal{R_X}_{23}(t) = \mathcal{R_X}_{32}(t) = \frac{2 \cos ^2(2 \tau ) \left(-A_-+A_++\frac{2 B_+ \xi _c}{\alpha _+}\right)+\frac{2 B_- \xi _c (\cos (4 \tau )-3)}{\alpha _-}}{8
   \left(A_-+A_+\right)},
\end{equation}

\small
\begin{equation}
    \mathcal{R_X}_{24}(t) =\mathcal{R_X}_{42}(t) = -\frac{\cos (2 \tau ) \left(\sin (2 \tau ) \left(A_--A_+-\frac{2 B_- \xi _c}{\alpha _-}-\frac{2 B_+ \xi _c}{\alpha _+}\right)+\frac{B_-
   \left(\xi _2-\xi _1\right)}{\alpha _-}-\frac{B_+ \left(\xi _1+\xi _2\right)}{\alpha _+}\right)}{4 \left(A_-+A_+\right)},
\end{equation}

\small
\begin{equation}
    \mathcal{R_X}_{33}(t) = \frac{\alpha _+ \left(-\left(\alpha _- A_- (\cos (4 \tau )-3)\right)+\alpha _- A_+ (\cos (4 \tau )+1)+4 B_- \left(\xi _1-\xi _2\right) \sin (2
   \tau )\right)+4 \left(\alpha _- B_++\alpha _+ B_-\right) \xi _c \cos ^2(2 \tau )}{8 \alpha _- \alpha _+ \left(A_-+A_+\right)},
\end{equation}

\small
\begin{equation}
    \mathcal{R_X}_{34}(t) = \mathcal{R_X}_{43}(t) = -\frac{\cos (2 \tau ) \left(\sin (2 \tau ) \left(A_--A_+-\frac{2 B_- \xi _c}{\alpha _-}-\frac{2 B_+ \xi _c}{\alpha _+}\right)+\frac{B_-
   \left(\xi _1-\xi _2\right)}{\alpha _-}-\frac{B_+ \left(\xi _1+\xi _2\right)}{\alpha _+}\right)}{4 \left(A_-+A_+\right)}.
\end{equation}

\section{Closed-form expressions for performance metrics of SQB}
\label{appC}
The assessment of the final state of the battery focuses on its ergotropy, which is quantified as the maximum work extractable in a cyclic unitary process. We calculate the ergotropy of our system based on X-gate charging as follows:

\begin{align}
\mathcal{E}=\frac{\sin ^2(\tau ) \left[4 \xi_c \left(\xi_c \cos (2 \tau ) \left(\frac{B_-}{\alpha_-}+\frac{B_+}{\alpha_+}\right)+\cos ^2(\tau ) \left(A_+- +A_-\right)\right)+\alpha _- B_-+\alpha _+ B_+\right]}{A_- +A_+}.
\end{align}
On the other hand, the instantaneous power $\mathcal{P}$ of the SQB is calculated as:
 \begin{align}
\mathcal{P} =\frac{\sin (2 \tau ) \left[\alpha _- \left\{4 \alpha _+ \left(A_+-A_-\right) \xi _c \cos (2 \tau )+B_+ \left(8 \xi _c^2 \cos (2 \tau )+\left(\xi
   _1+\xi _2\right){}^2\right)\right\}+\alpha _+ B_- \left(8 \xi _c^2 \cos (2 \tau )+\left(\xi _1-\xi _2\right){}^2\right)\right]}{\alpha _-
   \alpha _+ \left(A_-+A_+\right)}.
 \end{align}
Finally, the capacity $\mathcal{K}$ of the SQB is computed as:
 \begin{align}
\mathcal{K}=\frac{2 \left(A_-+A_+\right) \xi _c+\alpha _- B_-+\alpha _+ B_+}{2 \left(A_-+A_+\right)}.
 \end{align}

\end{widetext}

 \section{Parameter analysis}\label{appD}

In this appendix, we provide a detailed discussion of the physical meaning and selection basis of the parameters used in our simulations to ensure clarity and reproducibility.

\subsection{Selection of system parameters}

The model considered in this paper involves several key parameters: \(\xi_1\), \(\xi_2\), \(\xi_c\), and \(T\), which play crucial roles in defining the system's behavior.

\paragraph{Interaction parameters \(\xi_1\) and \(\xi_2\).}
The parameters \(\xi_1\) and \(\xi_2\) represent the Josephson energies of the qubits in the quantum battery system, directly influencing energy transfer and coherence dynamics. Their selection is based on typical values observed in superconducting qubit setups, ensuring that the studied regime is experimentally feasible. Specifically:
\begin{itemize}
    \item In Fig. \ref{figure1}, \(\xi_1\) is fixed at $1.5$ while \(\xi_2\) is varied over the values 0.1, 0.5, 1, and 2. This range allows us to explore different regimes, from weak ($\xi_2 = 0.1$) to strong ($\xi_2 = 2$) Josephson energy. At lower values, the qubit exhibits slower charging dynamics and weaker coherence effects, whereas at higher values, it contributes more significantly to energy storage and coherence preservation.
    \item In Fig. \ref{figure2}, \(\xi_2\) is fixed at 1.5 while \(\xi_1\) is varied over 0.1, 0.5, 1, and 2. This complementary analysis highlights the role of \(\xi_1\) in the system’s performance. Similar to \(\xi_2\), increasing \(\xi_1\) enhances energy storage efficiency and coherence retention, while lower values correspond to weaker qubit contributions to the charging process.
    \item The choice of \(\xi_1, \xi_2 = 1.5\) as a reference point, is motivated by previous theoretical studies \cite{Zidan_2023,POURKARIMI202327,benzahra2023quantum,chlih2024nonclassicality}, where similar values were found to optimize and amplify the quantum correlations and quantum coherence.
\end{itemize}

\paragraph{Coupling strength \(\xi_c\).}
The parameter \(\xi_c\) represents the coupling strength between the battery and its external driving mechanism. It is crucial in determining the rate of energy transfer.
\begin{itemize}
\item In Fig. \ref{figure3}, we analyze the impact of varying \(\xi_c\) while keeping \(\xi_1 = \xi_2 = 1.5\) fixed, allowing us to isolate its effect on energy dynamics. Specifically, we consider \(\xi_c = 0.1, 0.5, 1\), and \(2\), covering a range from weak to strong coupling, which enables us to assess how increasing interaction strength influences coherence and power transfer.
\item In Fig. \ref{figure4}, we repeat this analysis with a different parameter set (\(\xi_1 = 1.5, \xi_2 = 0.5\)) to confirm robustness across different regimes. Again, we use \(\xi_c = 0.1, 0.5, 1\), and \(2\), ensuring a broad exploration of interaction effects.
\item The choice of \(\xi_c = 0.05\) or \(\xi_c = 0.5\) in fixed cases (Figs. \ref{figure1} and \ref{figure2}) aligns with typical weak and moderate coupling regimes. 
\end{itemize}

\paragraph{Temperature \(T\).}
The temperature \(T\) affects the thermal fluctuations within the system. Our simulations explore two representative cases:
\begin{itemize}
\item  \(T = 0.5\) in Fig. \ref{figure1}, representing a higher temperature scenario where thermal effects might influence performance.
\item  \(T = 0.1\) in Figs. \ref{figure2}–\ref{figure4}, representing a lower temperature regime, consistent with dilution refrigerator environments used in superconducting qubits.
\end{itemize}

\subsection{Reproducibility considerations}

The parameter values were chosen to reflect realistic conditions for quantum battery implementations, ensuring that the numerical results are both theoretically grounded and experimentally feasible. To enhance reproducibility, we explicitly state the parameter values in all figure captions. Our analysis focuses on comparing distinct scenarios, such as weak vs. moderate coupling or low vs. high temperature, to illustrate the influence of key parameters on quantum battery performance. This approach allows for a clearer interpretation of how different system parameters affect energy storage and transfer.

\section*{Disclosures}
The authors declare that they have no known competing financial interests.

\section*{Data availability}
No datasets were generated or analyzed during the current study.

\bibliography{bibliography}
\end{document}